\documentclass[sigconf]{acmart}
\AtBeginDocument{%
  }

\usepackage{amsmath}
\usepackage{amsthm}
\usepackage{cancel}
\usepackage{mathtools}
\usepackage{graphicx}
\usepackage{subfigure}
\usepackage{siunitx}
\usepackage[inline]{enumitem}
\usepackage{geometry}
\usepackage{booktabs}
\usepackage{multirow}
\usepackage{acronym}
\usepackage{caption}
\usepackage{bm}

\usepackage{float}

\usepackage[capitalize]{cleveref}
\usepackage{multicol}
\usepackage{multirow}
\usepackage{makecell}

\usepackage{pifont}
\usepackage[table]{xcolor}

\copyrightyear{2026}
\acmYear{2026}
\setcopyright{cc}
\setcctype{by}
\acmConference[CIKM '26]{Proceedings of the 35th ACM International Conference on Information and Knowledge Management}{November 07--11, 2026}{Rome, Italy}
\acmBooktitle{Proceedings of the 35th ACM International Conference on Information and Knowledge Management (CIKM '26), November 07--11, 2026, Rome, Italy}
\acmDOI{10.1145/3799682.3840833}
\acmISBN{979-8-4007-2539-5/2026/11}

\newcommand{\tca}{\textsc{TCA}}
\newcommand{\tcagrpo}{\textsc{TCA-GRPO}}
\newcommand{\tcappo}{\textsc{TCA-PPO}}

\begin{document}

\title[Token-Level Credit Assignment Optimization for Generative Document Retrieval]{Token-Level Credit Assignment Optimization for \\ Generative Document Retrieval}

\author{Xinpeng Zhao}
\email{zhaoxp1001@gmail.com}
\authornote{Equal contribution.}
\affiliation{
  \institution{Shandong University}
  \city{Qingdao}
  \country{China}
}

\author{Yang Liu}
\authornotemark[1]
\email{202535775@mail.sdu.edu.cn}
\affiliation{
  \institution{Shandong University}
  \city{Jinan}
  \country{China}
}

\author{Ran Chen}
\email{chenran@stu.pku.edu.cn}
\affiliation{
  \institution{Peking University}
  \city{Beijing}
  \country{China}
}

\author{Xinyu Ma}
\email{xinyuma2016@gmail.com}
\affiliation{
  \institution{Baidu Inc.}
  \city{Beijing}
  \country{China}
}

\author{Daiting Shi}
\email{shidaiting01@baidu.com}
\affiliation{
  \institution{Baidu Inc.}
  \city{Beijing}
  \country{China}
}

\author{Pengjie Ren}
\email{jay.ren@outlook.com}
\affiliation{
  \institution{Shandong University}
  \city{Qingdao}
  \country{China}
}

\author{Zhumin Chen}
\email{chenzhumin@sdu.edu.cn}
\affiliation{
  \institution{Shandong University}
  \city{Qingdao}
  \country{China}
}

\author{Zhaochun Ren}
\authornote{Corresponding author.}
\email{z.ren@liacs.leidenuniv.nl}
\affiliation{
  \institution{Leiden University}
  \city{Leiden}
  \country{Netherlands}
}

\author{Xin Xin}
\authornotemark[2]
\email{xinxin@sdu.edu.cn}
\affiliation{
  \institution{Shandong University}
  \city{Qingdao}
  \country{China}
}
\renewcommand{\shortauthors}{Xinpeng Zhao et al.}

\begin{abstract}
Generative retrieval models perform document retrieval by autoregressively generating document identifiers (DocIDs). This process naturally forms a sequential decision problem, i.e., the model makes a sequence of token-level decisions, selecting a DocID token at each decoding step, with the resulting complete sequence identifying the retrieved document. However, relevance feedback is available only after the complete DocID has been generated and mapped to a document, resulting in a granularity mismatch between token-level generation decisions and document-level retrieval supervision. 
Consequently, existing reinforcement learning methods for generative retrieval rely on sequence-level rewards, assigning the same document-level relevance signal to every decoding step. Such uniform credit assignment obscures the contribution of individual token decisions, making it difficult to identify which decisions contribute to retrieval success or failure.

In this paper, we propose Token-Level Credit Assignment for Generative Retrieval (TCA), a fine-grained reinforcement learning framework that aligns the granularity of credit assignment with that of autoregressive DocID generation.
Unlike assigning a single reward to an entire generated DocID, TCA derives fine-grained rewards by comparing the hidden-state trajectory of each generated DocID with the gold DocID trajectory obtained from a frozen reference model.
These trajectory-based rewards provide differentiated feedback across decoding steps, allowing the policy to reinforce generation paths that remain aligned with the target DocID. 
Moreover, TCA decouples token-level credit assignment from policy optimization and can be instantiated with both GRPO and PPO. 
Experiments on MS MARCO and NQ benchmarks show that our method consistently outperforms baselines, demonstrating the effectiveness of fine-grained supervision for aligning autoregressive DocID generation with retrieval objectives.
\footnote{Code is released at \href{https://github.com/colinzhaoxp/TCA-RL4GR}{https://github.com/colinzhaoxp/TCA-RL4GR}.}
\end{abstract}



\maketitle

\begin{figure}[h]
    \centering
    \includegraphics[width=0.92\linewidth]{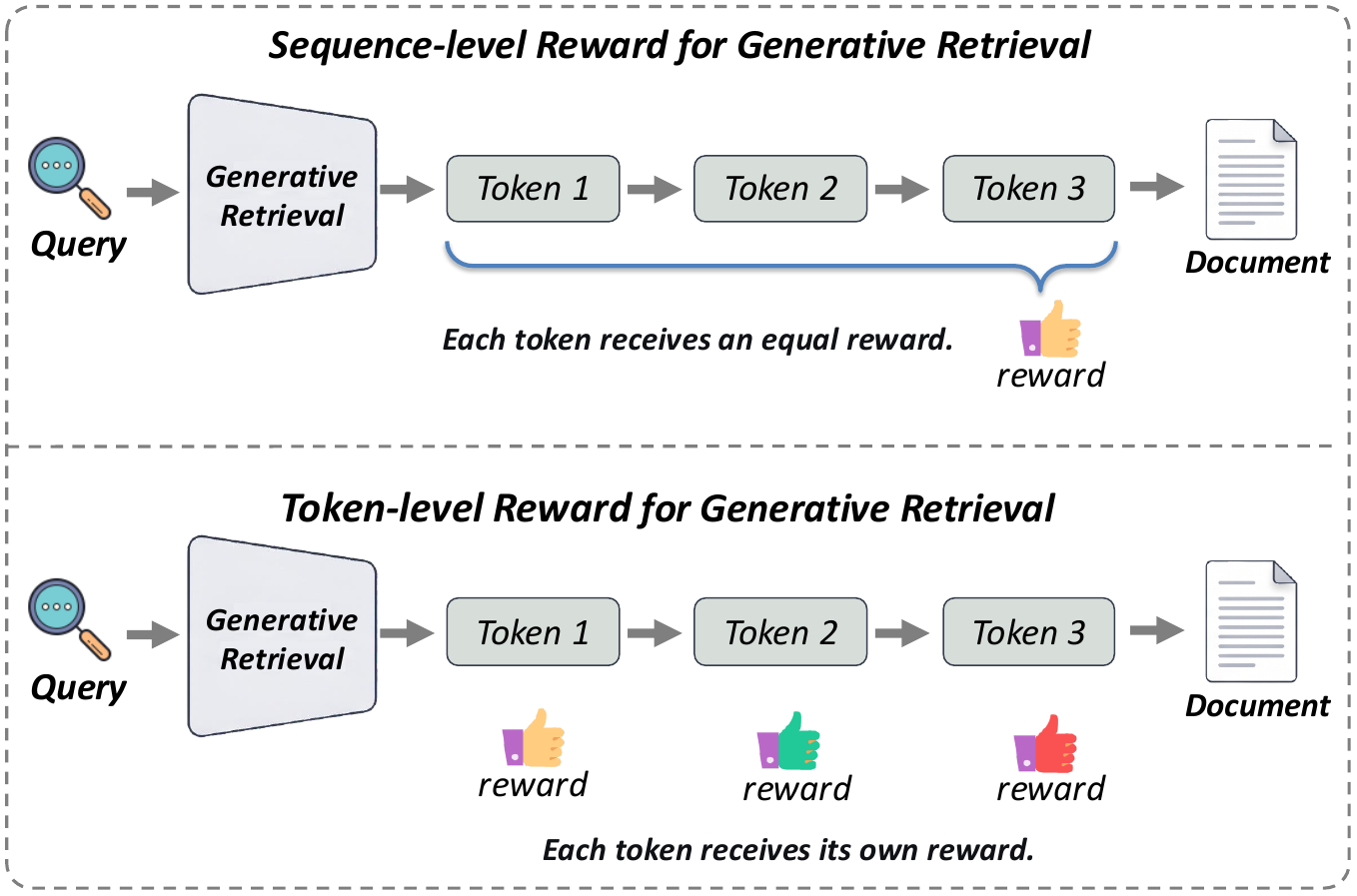}
    \caption{Comparison between sequence-level and token-level credit assignment in generative retrieval. Sequence-level rewards assign the same document-level signal to all generated DocID tokens, whereas token-level rewards provide step-wise feedback for individual token decisions.}
    \label{fig:reward-comparison}
    \vspace{-2mm}
\end{figure}

\section{Introduction}
Information retrieval is a fundamental component of modern information systems, supporting applications such as web search, open-domain question answering, and retrieval-augmented generation~\cite{zhu2025large, li2025matching}. 
Existing neural retrievers typically encode queries and documents into dense vectors, retrieve candidates through maximum inner-product search~\cite{Hofsttter2021EfficientlyTA, Izacard2021UnsupervisedDI}, and optionally apply a separate reranking stage. 
Although effective, this paradigm relies on an external vector index, fixed-dimensional vector representations, and multi-stage objectives that do not directly optimize the entire retrieval process for final relevance~\cite{DeCao2020AutoregressiveER, metzler2021rethinking, tay2022transformer}. 
Generative retrieval (GR) provides an alternative by directly generating the document identifier (DocID) of a relevant document conditioned on a query~\cite{tay2022transformer, NCI, GenRRL, DDRO, sun2026zerogr, zhang2026model}. 
By integrating corpus indexing and query-time retrieval within a unified sequence-to-sequence model, generative retrieval has demonstrated promising effectiveness and become an active research direction in neural information retrieval~\cite{zhang2026model,sun2026zerogr}.

Despite this progress, the training objective of generative retrieval remains imperfectly aligned with the goal of document retrieval~\cite{GenRRL, DDRO}. 
Most existing GR methods rely on autoregressive supervised fine-tuning, which maximizes the likelihood of each target DocID token~\cite{tay2022transformer, NCI, Ultron}. 
While this objective teaches the structure of the DocID space and the mapping from queries to identifiers, it does not directly optimize retrieval effectiveness. 
Specifically, DocID generation proceeds through a sequence of token-level decisions, whereas relevance can be assessed only after the complete DocID has been generated. This creates a granularity mismatch between token-level optimization and document-level retrieval judgment.
A generated DocID may contain several locally plausible tokens yet ultimately refer to an irrelevant document; moreover, an intermediate token may be important because it routes the subsequent decoding process toward a region from which relevant DocIDs remain reachable.
Standard next-token likelihood training does not explicitly distinguish these effects, as it optimizes the imitation of observed DocIDs rather than the relevance of documents retrieved through the model's own generation trajectories.

Reinforcement learning provides a principled way to bridge this gap by treating DocID generation as a decision-making process and optimizing the generation policy with relevance feedback~\cite{RLHF}. 
Recent studies have applied relevance-based reinforcement learning~\cite{GenRRL} and preference-based optimization~\cite{DDRO} to better align generated DocIDs with document-level relevance.
However, the relevance score becomes available only after the complete DocID has been generated, and the resulting document-level signal is assigned uniformly to all tokens. 
Such uniform credit assignment fails to capture the distinct contributions of individual token decisions~\cite{CA}. 
During autoregressive DocID generation, different tokens play different roles: Early tokens sharply constrain the set of reachable documents, intermediate tokens refine semantic or structural distinctions, and later tokens identify the exact document. 
Uniform rewards therefore obscure the effects of individual token decisions on retrieval outcomes.
Motivated by this, we argue that effective reinforcement learning for generative retrieval requires token-level credit assignment. We formulate the DocID generation as a sequential decision problem. 
At each decoding step, the state consists of the query and the partially generated DocID, whereas the action is the next DocID token, and the complete action sequence determines the retrieved document. 
Under this formulation, the key challenge is to attribute the final retrieval outcome to the token decisions made along the generation trajectory. 
A fine-grained reward should distinguish decisions that keep relevant DocIDs reachable from those that steer subsequent decoding away from them, rather than assigning the same terminal signal to every step.

To address this challenge, we propose \emph{Token-Level Credit Assignment for Generative Retrieval} (\textsc{TCA}), a fine-grained reinforcement learning framework that separates token-level reward design from the choice of policy optimization algorithm. 
The training of TCA consists of two stages: First, we perform supervised fine-tuning to endow the model with basic query-to-DocID generation ability. Second, we refine the model with reinforcement learning under token-level credit assignment. 
Instead of assigning one reward to the entire generated sequence, \textsc{TCA} provides fine-grained feedback for intermediate DocID tokens by comparing the generated hidden-state trajectory with the gold DocID trajectory from a frozen reference model. 
This trajectory-based reward gives the policy a dense signal for distinguishing helpful and harmful token decisions. 
As an optimizer-agnostic credit assignment framework, \textsc{TCA} assigns fine-grained credit to individual DocID tokens while remaining compatible with different reinforcement learning algorithms. In this paper, we consider two representative implementations: \textsc{TCA-GRPO}, which uses group-relative advantage estimation to compare multiple candidate DocIDs for the same query, and \textsc{TCA-PPO}, which uses value-function-based advantage estimation under the same token-level reward design. Both implementations are compatible with different forms of DocIDs and can be combined with constrained decoding over the legal DocID space.

We conducted experiments on the MS MARCO(MS) and Natural Questions(NQ) benchmarks using both title--URL (TU) and product-quanti\-zation (PQ) DocIDs to evaluate the effectiveness and generality of token-level credit assignment.  
Both \tcagrpo{} and \tcappo{} improve $R@1$ and $MRR@10$. 
Compared with the DDRO(MS-TU) and DDRO(NQ-PQ), \textsc{TCA-GRPO} achieves relative improvements of 3.56\%/1.53\% on MS-TU and 3.99\%/3.28\% on NQ-PQ in $R@1$/$MRR@10$, respectively. The results show that \tca{} consistently improves generative retrieval performance over supervised fine-tuning and sequence-level reinforcement learning baselines, while the comparison between GRPO and PPO further shows how different advantage estimation strategies affect token-level reinforcement learning for DocID generation. We further analyze different reward designs and training components, demonstrating that fine-grained token rewards provide a robust signal for aligning autoregressive DocID generation with document-level relevance.

The main contributions of this work are summarized as follows:
\begin{enumerate*}[label=(\arabic*),leftmargin=*]
    \item We formulate generative document retrieval as a sequential decision problem and identify fine-grained credit assignment as a central limitation of existing reinforcement learning methods based on sequence-level rewards.
    \item We propose \textsc{TCA}, a fine-grained reinforcement learning framework that assigns relevance-aware rewards to individual DocID tokens and can be instantiated with both GRPO and PPO for policy optimization.
    \item We conduct extensive experiments and analyses on standard retrieval benchmarks, showing that token-level credit assignment improves retrieval effectiveness over strong supervised and reinforcement learning baselines.
\end{enumerate*}

\section{Related Work}


\subsection{Generative retrieval}
Generative retrieval (GR) formulates document retrieval as an autoregressive generation problem, where a model directly generates document identifiers (DocIDs) for a given query. Existing studies have mainly advanced this paradigm along four dimensions:
(i) \textit{Identifier construction}: Existing DocIDs can be broadly divided into lexical and numerical. Lexical identifiers, such as titles, URLs~\citep{DeCao2020AutoregressiveER,Chen2022CorpusBrainPA,Ultron}, representative n-grams, or summaries~\citep{Bevilacqua2022AutoregressiveSE,Li2023SummarizationBasedDI}, are human-readable and easy to interpret. Numerical identifiers instead encode documents as discrete semantic codes, including clustering-based semantic codes~\citep{tay2022transformer} and RQ-VAE-based quantized identifiers~\citep{Wang2024ContentBasedCG,zeng2023scalable_and_effective,NOVO}.
(ii) \textit{Index construction}: Existing methods enhance indexing construction through document chunking~\citep{tay2022transformer}, pseudo-query generation~\citep{Zhuang2022BridgingTG}, and rehearsal-based augmentation~\citep{Tang2023SemanticEnhancedDS}. Other studies further investigate multi-granular synthetic data generation~\citep{Wen2025OnSD} and continual indexing for dynamic corpora~\citep{DSI++,Chen2023ContinualLF,zhang2025replication,kidist2026parametric,zhang2026model, tang2025generative}.
(iii) \textit{Decoding strategy}: Existing methods can be broadly divided into autoregressive and non-autoregressive paradigms. Autoregressive decoding generates DocIDs token by token, typically using constrained beam search to restrict the output space to valid DocIDs~\citep{DeCao2020AutoregressiveER,tay2022transformer, Ren2023TOMEAT,Li2023MultiviewIE, zeng2024planning,kidist2026lost,chen2026closing}. 
Recent studies further explore non-autoregressive DocID generation to improve efficiency and global control~\citep{valluri2024scaling, zhao2025diffugr}. These methods generate multiple DocID tokens in parallel or refine identifiers through iterative procedures, such as diffusion-based language models and parallel generation frameworks.
(iv) \textit{Training objective}: Early generative retrieval models are mainly optimized by supervised maximum-likelihood training~\citep{tay2022transformer,NCI,DeCao2020AutoregressiveER}. 
Later work incorporates ranking-oriented supervision, including marginal ranking loss~\citep{li2024learning}, knowledge distillation~\citep{li2024distillation}, and contrastive learning~\citep{cheng2025descriptive,DOGR}.
Recent RL-based methods further leverage retrieval feedback for optimization: GenRRL~\citep{GenRRL} adopts reinforcement learning from relevance feedback with an external reward model, while DDRO~\citep{DDRO} leverages pairwise relevance feedback without explicit reward modeling.
In contrast, our work studies GR from a sequential decision-making perspective and views DocID generation as a trajectory of token-level decisions, where each generated token affects the final retrieval quality. 

\subsection{Credit assignment problem}
Credit assignment concerns how a training signal should be distributed to the individual decisions that produced it. 
In reinforcement learning for LLMs, the central challenge is often token-level: although generation proceeds autoregressively, supervision in RLHF is typically provided only as a sequence-level reward for the final response~\citep{RLHF}. 
Some methods introduce denser supervision directly, such as sentence-level human feedback~\citep{wu2023fine}, while a larger body of work infers token-level rewards from response-level preference signals through reward densification~\citep{chan2024dense}, reward redistribution~\citep{li2025red}, continuous token-level reward modeling~\citep{yoon2024tlcr}, and token-wise preference optimization objectives~\citep{zhong2024dpo, zeng2024token, yang2025selective}. 
Although these studies demonstrate the importance of fine-grained credit assignment, they are not directly tailored to generative retrieval. 
Unlike open-ended text generation, generative retrieval produces a document identifier as a constrained token sequence, in which each token can be viewed as a routing decision over the document space. 
Therefore, credit assignment in generative retrieval is more sensitive to structural dependencies: the model needs to understand how each token decision contributes to identifying the correct document, rather than simply improving the overall quality of a generated response. 
A closely related method is TPMA-GRPO in OneSearch-V2 \cite{OneSearch-V2}, which addresses uniform sequence-level credit assignment by deriving position-level marginal rewards from prefix matching and gating downstream gradients according to prefix correctness. In contrast, our TCA assigns continuous token-level rewards by measuring hidden-state trajectory similarity between generated and gold DocIDs, without relying on an explicit hierarchical identifier structure or exact prefix matching. 

\begin{figure*}[t]
    \centering
    \includegraphics[width=0.92\linewidth]{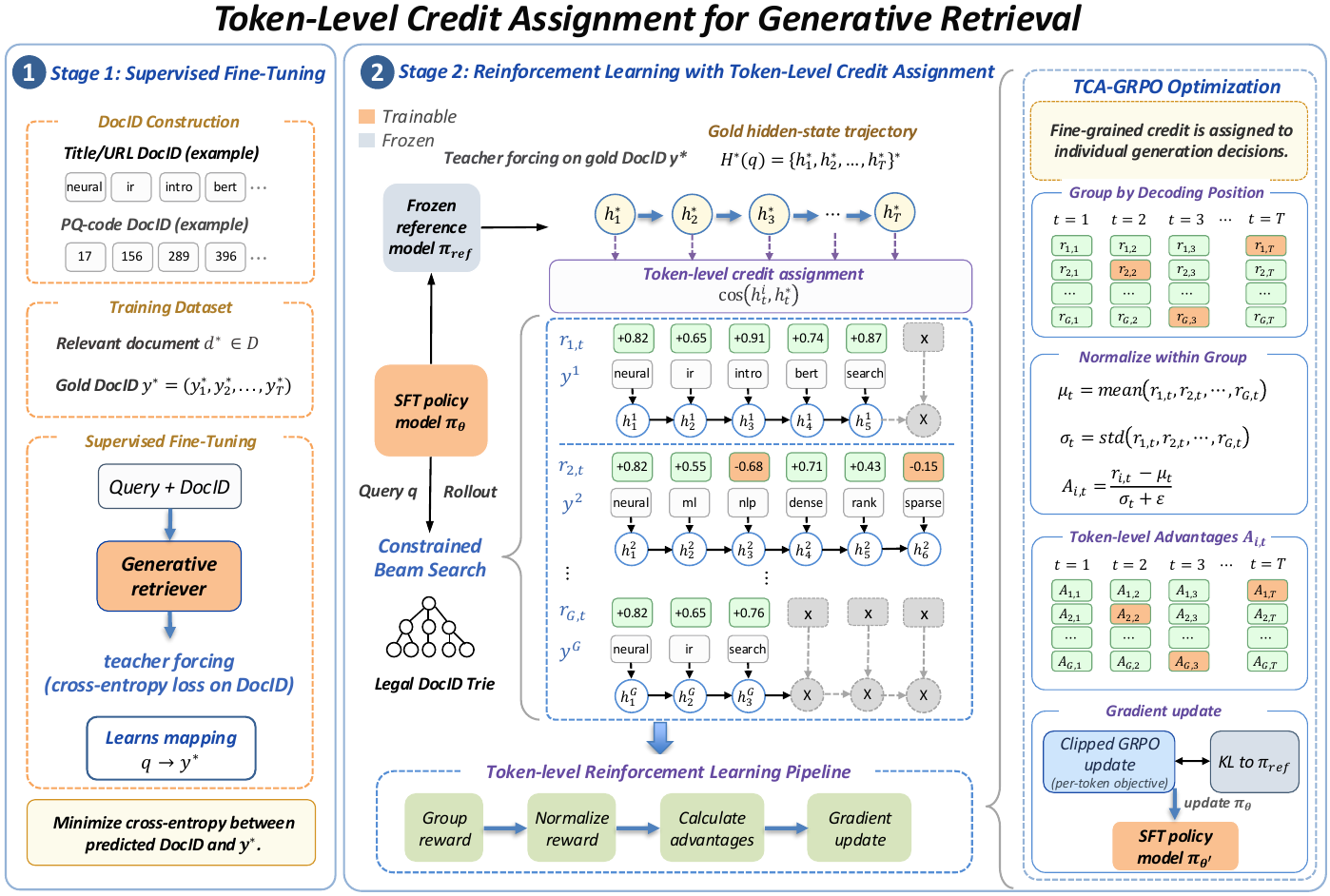}
    \caption{Overview of the proposed Token-Level Credit Assignment for Generative Retrieval with GRPO (TCA-GRPO).}
    \label{fig:pipeline}
\end{figure*}

\section{Preliminary}

\subsection{Generative retrieval}
Generative information retrieval formulates document retrieval as a sequence generation problem. Given a query $q$, a generative retriever directly generates a document identifier (DocID) sequence:
$y=(y_1,y_2,\ldots,y_n)$,
where $y_t$ denotes the token generated at decoding step $t$, and $n$ denotes the length of the current DocID sequence only for notational convenience; different documents may have DocIDs with different lengths. The autoregressive generation probability is factorized as:
\begin{equation}
    \pi_\theta(y|q)
    =
    \prod_{t=1}^{n}
    \pi_\theta(y_t|q,y_{<t}),
\end{equation}
where $\pi_\theta$ denotes the generative retrieval model parameterized by $\theta$, and $y_{<t}=(y_1,\ldots,y_{t-1})$ is the prefix generated before step $t$.
During retrieval, the model generates a set of candidate DocIDs and ranks them according to their generation probabilities. The ranking score of a candidate DocID is written as:
\begin{equation}
    S_\theta(y|q)
    =
    \log \pi_\theta(y|q)
    =
    \sum_{t=1}^{n}
    \log \pi_\theta(y_t|q,y_{<t}).
\end{equation}
Thus, unlike dense retrieval which typically computes a ranking score using a single query-document similarity function, the score in generative retrieval decomposes into a sum of multiple token-level generation probabilities. Consequently, each token decision can affect the probability of the complete DocID.

\subsection{Generative retrieval as sequential decision problem}
\label{sec:prelim-seq-decision}

The autoregressive generation of DocID naturally defines a multi-step decision-making process. 
At decoding step $t$, the state is defined as $s_t=(q,y_{<t})$ which consists of the query and the generated DocID prefix,
whereas the action is the next DocID token $a_t=y_t$.
The state transition is $s_{t+1}=(q,y_{\leq t})$.
In practice, DocID decoding is constrained to valid identifiers associated with documents in the corpus. 

This formulation highlights a key credit assignment issue in generative retrieval. The final retrieval outcome is determined by a complete DocID sequence, but each intermediate token decision affects the future prefix space and the final ranking score. Existing document-level or sequence-level objectives provide limited feedback for these intermediate decisions. This motivates the token-level reinforcement learning objective developed next.

\section{Method}
\subsection{Overview}

We propose \emph{Token-Level Credit Assignment for Generative Retrieval} (\textsc{TCA}), a reinforcement learning framework that assigns credit to individual DocID tokens in generative document retrieval. Given a query, a generative retriever produces a document identifier token by token, and the generated identifier is mapped to a document in the corpus. 
Instead of assigning a single sequence-level reward uniformly to all generated tokens, our key idea is to estimate token-level rewards for intermediate decoding steps and use them to update the generation policy.

As shown in Figure~\ref{fig:pipeline}, the training pipeline contains two stages. First, supervised fine-tuning (SFT) initializes the model with query-to-DocID generation ability. 
Second, the model is further refined through reinforcement-learning post-training with token-level credit assignment. 
In this stage, an SFT model is frozen as a reference model, gold DocID trajectories are cached under teacher forcing, and the current policy generates candidate DocIDs under constrained decoding. Token-level rewards are then computed by comparing the generated hidden-state trajectory with the gold trajectory, and group relative policy optimization is used to update the policy. 
This design combines token-level credit assignment with the stability of group-relative advantage estimation.

\subsection{Training problem formulation}
We specify the training problem under the sequential decision view introduced in Section~\ref{sec:prelim-seq-decision}. Let $\mathcal{D}$ denote the document corpus. Each training example consists of a query $q$, a ground-truth relevant document $d^\ast \in \mathcal{D}$, and its corresponding gold DocID:
\begin{equation}
    y^\ast=(y_1^\ast,y_2^\ast,\ldots,y_T^\ast),
\end{equation}
where $T$ denotes the length of the corresponding gold DocID tokens and the superscript $\ast$ indicates the ground truth.
During rollout, the policy $\pi_\theta$ autoregressively generates a candidate DocID:
\begin{equation}
    \hat{y}=(y_1,y_2,\ldots,y_{n}).
\end{equation}
A complete generated DocID is mapped to a document $d_{\hat{y}}$ through the corpus DocID table. Retrieval quality is therefore determined by the relevance of $d_{\hat{y}}$ to $q$, while the model is updated through the token-level probabilities $\pi_\theta(y_t \mid q,y_{<t})$.

The central problem is to bridge this granularity mismatch. The supervision available in training identifies the relevant document and its gold DocID, but the policy decisions are made one token at a time along a generated trajectory. Our goal is to convert the document-level target and the gold generation trajectory into token-level rewards $\{r_{i,t}\}_{t=1}^{n}$ and token-level advantages $\{A_{i,t}\}_{t=1}^{n}$. These signals should distinguish helpful and harmful intermediate decisions within the same DocID trajectory, so that reinforcement learning can improve retrieval behavior at the granularity of the actual generation process.

\subsection{Supervised fine-tuning for generative retrieval}

We first initialize a base generative retriever through standard supervised fine-tuning. Here we consider two representative DocID designs. 
Title--URL (TU) uses document titles and URLs as identifiers to preserve lexical and self-descriptive document information, whereas product-quantization (PQ) identifiers discretize document representations into semantic code tokens through vector quantization.

Given a query--DocID pair $(q,y^\ast)$, the SFT objective maximizes the likelihood of the gold DocID under teacher forcing:
\begin{equation}
    \mathcal{L}_{\mathrm{SFT}}
    =
    -\sum_{t=1}^{T}
    \log \pi_\theta(y_t^\ast \mid q,y_{<t}^\ast).
\end{equation}
This objective teaches the model the syntax of the DocID space and the basic alignment between queries and document identifiers. It also provides a stable initialization for reinforcement learning, which is important because unconstrained exploration over the full vocabulary can easily produce invalid or low-quality DocID sequences.

However, SFT only imitates the observed gold tokens and does not directly optimize document-level retrieval relevance. A token that receives high likelihood under teacher forcing may still lead to an irrelevant document when generated under the model's own prefix. Conversely, an early token decision may be valuable because it routes the trajectory toward a promising region of the DocID space, even though its contribution is not visible from the local likelihood term. This limitation motivates the second training stage, where we use relevance-aware token-level rewards.

\subsection{Token-level credit assignment}

The goal of token-level credit assignment is to provide position-specific feedback for each DocID generation decision. 
Rather than assigning a single scalar reward to the whole generated identifier, we compare the hidden-state trajectory of a generated DocID with that of the gold DocID.

\noindent \textbf{Gold trajectory cache.} For each training pair $(q,y^\ast)$, we first freeze the SFT model as the reference model and construct a hidden-state trajectory for the gold DocID. 
At the $t$-th decoding step, the reference model is conditioned on the query $q$ and the gold prefix $y_{<t}^\ast$ under teacher forcing. 
We retain the final-layer decoder hidden state $h_t^\ast$ prior to the LM head, which is used to predict the gold token $y_t^\ast$.
The gold trajectory cache is computed once before reinforcement learning and remains fixed throughout policy optimization.

\noindent\textbf{Generated trajectory.} During reinforcement learning, the current policy generates a candidate DocID $y^i=(y_1^i,\ldots,y_{n}^i)$ under trie-constrained decoding. 
At the $t$-th decoding step, the policy is conditioned on $q$ and the generated prefix $y_{<t}^i$, and we retain the  final-layer decoder hidden state $h_t^i$ prior to the LM head. 
Padding positions are masked out and do not contribute to either reward or policy update.

\noindent\textbf{Trajectory reward.} At each valid decoding position covered by both trajectories, we compute the trajectory reward by measuring the cosine similarity between the corresponding generated and gold hidden states:
\begin{equation}
    r_{i,t}^{\mathrm{traj}}
    =
    \begin{cases}
        \cos(h_t^i, h_t^\ast), & t \le T, \\
        0, & t > T.
    \end{cases}
\end{equation}
When the generated DocID extends beyond the gold DocID, no position-aligned gold state is available. We therefore apply an overflow penalty to each additional token:
\begin{equation}
    r_{i,t}^{\mathrm{len}}
    = 
    \begin{cases}
        0, & t \le T, \\
        -0.5, & t > T.
    \end{cases}
\end{equation}

Although the trajectory reward provides position-specific feedback, it does not explicitly capture whether the completed DocID identifies the target document. We therefore introduce an exact-match bonus $R_{\mathrm{hit}}=2$, which is assigned to every valid token position when the generated DocID matches the gold DocID. Combining the trajectory-alignment reward, length penalty, and exact-match bonus, the final token-level reward is as follows:
\begin{equation}
    r_{i,t}
    = r_{i,t}^{\mathrm{traj}}
    + r_{i,t}^{\mathrm{len}}
    + \mathbf{1}[y^i=y^\ast]R_{\mathrm{hit}},
\end{equation}
where $\mathbf{1}[\cdot]$ denotes the indicator function.
The exact-match bonus is applied to all valid positions of an exactly matched DocID rather than only to the final token.

In this reward design, the trajectory reward provides position-specific feedback, the length penalty discourages over-generation, and the exact-match bonus maintains a direct connection to the final retrieval target.
This reward construction does not require a separately trained reward model.

\subsection{\textsc{TCA-GRPO} optimization}

\textsc{TCA-GRPO} converts token-level rewards into group-relative advantages for policy optimization~\cite{shao2024deepseekmath}. For each query $q$, the policy generates a group of $G$ candidate DocIDs:
\begin{equation}
    \mathcal{G}(q)=\{y^{1},y^{2},\ldots,y^{G}\}.
\end{equation}
In our implementation, GRPO candidates are generated with constrained beam search. For each generated candidate, token rewards are computed at valid generated positions. Let $m_{i,t}$ denote whether position $t$ of candidate $y^i$ is a valid non-padding token position:
\begin{equation}
    m_{i,t}
    =
    \begin{cases}
    1, & y^i_t \neq \mathrm{pad\ token}, \\
    0, & y^i_t = \mathrm{pad\ token}.
    \end{cases}
\end{equation}

At each decoding position, rewards are normalized across the candidate group. The position-wise reward mean and variance are:
\begin{equation}
    \mu_t =
    \frac{\sum_{i=1}^{G} m_{i,t} r_{i,t}}
    {\sum_{i=1}^{G} m_{i,t}},
    \quad
    \sigma_t^2 =
    \frac{\sum_{i=1}^{G} m_{i,t}(r_{i,t}-\mu_t)^2}
    {\sum_{i=1}^{G} m_{i,t}}.
\end{equation}
For each valid token position, the token-level GRPO advantage is
\begin{equation}
    A_{i,t}^{\mathrm{GRPO}}
    =
    \frac{r_{i,t}-\mu_t}{\sigma_t+\epsilon},
\end{equation}
where $\epsilon$ is a small positive constant for numerical stability.

This position-wise normalization compares candidates at the same decoding depth and avoids learning a value function over discrete DocID prefix states.

Let $\pi_{\theta_{\mathrm{old}}}$ denote the policy used to generate the rollout candidates. For each generated token $y_t^i$, the probability ratio between the updated and rollout policies is
\begin{equation}
    \rho_{i,t}
    =
    \frac{
    \pi_\theta(y_t^{i} \mid q,y_{<t}^{i})
    }{
    \pi_{\theta_{\mathrm{old}}}(y_t^{i} \mid q,y_{<t}^{i})
    }.
\end{equation}
The clipped token-level objective is
\begin{equation}
\begin{aligned}
    \mathcal{L}_{\mathrm{TCA\mbox{-}GRPO}}
    =
    & -\mathbb{E}_{i,t}
    \Big[
    m_{i,t}
    \min \Big(
    \rho_{i,t}A_{i,t}^{\mathrm{GRPO}},
    \\
    & \mathrm{clip}(\rho_{i,t},1-\epsilon_{\mathrm{clip}},1+\epsilon_{\mathrm{clip}})
    A_{i,t}^{\mathrm{GRPO}}
    \Big)
    \Big]
    +
    \beta\mathcal{L}_{\mathrm{KL}} .
\end{aligned}
\end{equation}

The KL term regularizes the updated policy toward the SFT reference model:
\begin{equation}
    \mathcal{L}_{\mathrm{KL}}
    =
    \mathbb{E}_{i,t}
    \left[
    m_{i,t}
    KL
    \left(
    \pi_\theta(\cdot \mid q,y_{<t}^{i})
    \|
    \pi_{\mathrm{ref}}(\cdot \mid q,y_{<t}^{i})
    \right)
    \right].
\end{equation}
In practice, this KL is estimated on generated tokens and is active only when $\beta>0$.

The key difference from sequence-level reinforcement learning lies in the granularity of the advantage signal. Sequence-level methods assign the same advantage to all tokens in a generated DocID, whereas \textsc{TCA-GRPO} computes a separate advantage at each decoding position. This allows the policy update to reinforce or weaken intermediate token decisions within the same DocID trajectory.

\subsection{\textsc{TCA-PPO} optimization} \label{sec:TCA-PPO}

For comparison, we implement a PPO-style actor-critic variant~\cite{PPO}, denoted \textsc{TCA-PPO}.
It uses the same token-level reward \(r_{i,t}\) as \textsc{TCA-GRPO}, but replaces group-relative advantage estimation with a learned value baseline.
A linear value head is attached to the pre-action decoder hidden state \(h_t^i\) (corresponding to state \(s_t^i=(q,y_{<t}^i)\)):
\begin{equation}
    V_\psi(s_t^i) = V_\psi(h_t^i),
\end{equation}
which estimates the expected future token-level return from the current DocID prefix.

When KL regularization is enabled, we apply it as reward shaping against the frozen SFT reference policy \(\pi_{\mathrm{ref}}\):
\begin{equation}
    \tilde r_{i,t}
    =
    r_{i,t}
    -
    \beta\bigl(
    \log \pi_{\theta_{\mathrm{old}}}(y_t^i\mid q,y_{<t}^i)
    -
    \log \pi_{\mathrm{ref}}(y_t^i\mid q,y_{<t}^i)
    \bigr).
\end{equation}
When \(\beta=0\), we simply have \(\tilde r_{i,t}=r_{i,t}\).
Token-level advantages are then computed from \(\tilde r_{i,t}\) using standard generalized advantage estimation (GAE)~\cite{schulman2015high} with discount \(\gamma=1.0\) and trace-decay \(\lambda\). Advantages are normalized to zero mean and unit variance over all valid positions in the mini-batch.

The clipped token-level policy loss follows standard PPO:
\begin{equation}
\begin{aligned}
    \mathcal{L}_{\mathrm{policy}}
    =
    -\mathbb{E}_{i,t}
    \Big[
    m_{i,t}
    \min \Big(
    &\rho_{i,t}A_{i,t}^{\mathrm{PPO}},\\
    &\mathrm{clip}(\rho_{i,t},1-\epsilon_{\mathrm{clip}},1+\epsilon_{\mathrm{clip}})
    A_{i,t}^{\mathrm{PPO}}
    \Big)
    \Big],
\end{aligned}
\end{equation}
where \(\rho_{i,t}=\pi_\theta(y_t^i\mid q,y_{<t}^i)/\pi_{\theta_{\mathrm{old}}}(y_t^i\mid q,y_{<t}^i)\) is the probability ratio and \(m_{i,t}\) masks out padding positions. The value head is trained with a mean-squared-error loss against GAE return targets \(R_{i,t}\):
\begin{equation}
    \mathcal{L}_{\mathrm{value}}
    =
    \frac{1}{2}\mathbb{E}_{i,t}\Big[m_{i,t}(V_\psi(s_t^i)-R_{i,t})^2\Big].
\end{equation}
The final objective is
\begin{equation}
    \mathcal{L}_{\mathrm{TCA\mbox{-}PPO}}
    =
    \mathcal{L}_{\mathrm{policy}}
    +
    c_v\mathcal{L}_{\mathrm{value}},
\end{equation}
where \(c_v=0.1\) is the value-loss coefficient.
We omit the entropy bonus because the trie-constrained action space and KL reward shaping already prevent policy collapse.
This actor-critic variant provides a value-based counterpart to \textsc{TCA-GRPO}: \textsc{TCA-PPO} estimates token-level advantages from a learned prefix-state value function, whereas \textsc{TCA-GRPO} obtains them through group-relative normalization at each decoding position.


\section{Experimental Setups}

\subsection{Datasets}
\textbf{Natural Questions}
Natural Questions(NQ)~\citep{Kwiatkowski2019NaturalQA} is a widely applied benchmark for evaluating retrieval models~\citep{tay2022transformer, NCI}. 
We use a subset NQ320k that consists of about 320k query-document pairs, where the documents are gathered from Wikipedia pages, and the queries are natural language questions.
Note that following the data processing pipeline of prior work~\citep{DDRO}, and we obtain approximately 100k documents.

\noindent\textbf{MS MARCO}
MS MARCO(MS)~\citep{Campos2016MSMA} document dataset is a collection of queries and web pages from Bing search.
Similarly to NQ320k dataset, we rank the documents of the MS MARCO dataset by the number of associated queries and keep the top about 320K documents and use their corresponding queries for training as described in prior works~\citep{Ultron, DDRO}. 
We evaluate the models on the queries of the MS MARCO document dev set and retrieval on the sampled document subset as described in prior works~\citep{DDRO}.

\subsection{Baselines and metrics}
We compare our method with representative term-based, dense, and generative retrieval baselines. For term-based retrieval, we include BM25 and DocT5Query. For dense retrieval methods, we include DPR~\cite{karpukhin-etal-2020-dense}, ANCE~\cite{Xiong2020ApproximateNN}, RepBERT~\cite{zhan2020repbert}, and Sentence-T5~\cite{Ni2021SentenceT5SS}. For generative retrieval, we include DSI~\cite{tay2022transformer}, DSI-QG~\cite{Zhuang2022BridgingTG}, NCI~\cite{NCI}, SEAL~\cite{Bevilacqua2022AutoregressiveSE}, Ultron~\cite{Ultron}, ROGER~\cite{ROGER}, MINDER~\cite{Li2023MultiviewIE}, and LTRGR\cite{li2024learning}. We also compare with GenRRL~\cite{GenRRL}, a reinforcement learning based generative retrieval method.
Following prior works, we evaluate retrieval performance using Recall@\{1,5,10\}, and Mean Reciprocal Rank at 10 (MRR@10), which jointly measure top-ranked retrieval accuracy and ranking quality within the top 10 results.

\textbf{Note on result sourcing.} Since the source code of GenRRL\cite{GenRRL} and ROGER \cite{ROGER} is not publicly accessible, their baseline performance is cited directly from the corresponding original papers. 
To enable a controlled evaluation, we independently reproduce other baseline methods under the same experimental setup and report the reproduced results by reproducing the released code using the dataset configurations adopted in this work.

\subsection{DocID settings}
We evaluate two types of DocIDs: Title+URL (TU) and Product Quantization (PQ) following DDRO~\cite{DDRO}. TU uses document titles and URLs as identifiers, preserving lexical information that is useful for web search scenarios. PQ represents documents with discrete semantic codes obtained from quantized document representations, which can better capture semantic information. 

\begin{table}[t]
\centering
\scriptsize
\caption{Main results on MS MARCO document ranking.
$\ddagger$ indicates a statistically significant improvement over the baseline under the same DocID type, as determined by a paired $t$-test at the $p < 0.05$ level. The best results are shown in bold, and the second-best results are underlined.}
\label{tab:msmarco_main}
\setlength{\tabcolsep}{6pt}
\resizebox{\columnwidth}{!}{
\begin{tabular}{l | cccc}
\toprule
Model & R@1 & R@5 & R@10 & MRR@10 \\
\midrule
\multicolumn{5}{l}{\textit{Term-based \& Dense retrieval}} \\
\midrule
BM25 & 18.94 & 42.82 & 55.07 & 29.24 \\
DocT5Query & 23.27 & 49.38 & 63.61 & 34.81 \\

DPR & 29.08 & 62.75 & 73.13 & 43.41 \\
ANCE & 29.65 & 63.43 & 74.28 & 44.09 \\
RepBERT & 25.25 & 58.41 & 69.18 & 38.48 \\
Sentence-T5 & 27.27 & 58.91 & 72.15 & 40.69 \\

\midrule
\multicolumn{5}{l}{\textit{Generative retrieval}} \\
\midrule
DSI (SI) & 25.74 & 43.58 & 53.84 & 33.92 \\
DSI-QG (SI) & 28.82 & 50.74 & 62.26 & 38.45 \\
NCI (SI) & 29.54 & 57.99 & 67.28 & 40.46 \\
SEAL (NG) & 27.58 & 52.47 & 61.01 & 37.68 \\
Ultron (TU) & 29.82 & 60.39 & 68.31 & 42.53 \\
Ultron (PQ) & 31.55 & 63.98 & 73.14 & 45.35 \\
ROGER-NCI (SI) & 30.61 & 59.02 & 68.78 & 42.02 \\
ROGER-Ultron (TU) & 33.07 & 63.93 & 75.13 & 46.35 \\
MINDER (SI) & 29.98 & 58.37 & 71.92 & 42.51 \\
LTRGR (SI) & 32.69 & 64.37 & 72.43 & 47.85 \\

\midrule
\multicolumn{5}{l}{\textit{RL-based Generative retrieval}} \\
\midrule
GenRRL (TU) & 33.01 & 63.62 & 74.91 & 45.93 \\
GenRRL (Sum) & 33.23 & 64.48 & \textbf{75.80} & 46.62 \\
DDRO (PQ) & 31.19 & 63.00 & 71.91 & 44.41 \\
DDRO (TU) &\underline{38.24} & \underline{67.33} &{74.26} &\underline{50.33} \\

\midrule
\multicolumn{5}{l}{\textit{Ours}} \\
\midrule
TCA-GRPO (PQ) & 32.42\rlap{$^\ddagger$} & 61.88 & 71.16 & 44.85 \\
TCA-GRPO (TU) & \textbf{39.60}\rlap{$^\ddagger$} & \textbf{68.19}\rlap{$^\ddagger$} & \underline{75.37} & \textbf{51.10}\rlap{$^\ddagger$} \\
\bottomrule
\end{tabular}
}
\end{table}
\begin{table}[t]
\centering
\scriptsize
\caption{Main results on Natural Questions document ranking. 
$\ddagger$ indicates a statistically significant improvement over the baseline under the same DocID type, as determined by a paired $t$-test at the $p < 0.05$ level. The best results are shown in bold, and the second-best results are underlined.}
\label{tab:nq_main}
\resizebox{\columnwidth}{!}{
\begin{tabular}{l | cccc}
\toprule
Model & R@1 & R@5 & R@10 & MRR@10 \\
\midrule
\multicolumn{5}{l}{\textit{Term-based \& Dense retrieval}} \\
\midrule
BM25 & 14.06 & 36.91 & 47.93 & 23.60 \\
DocT5Query & 19.07 & 43.88 & 55.83 & 29.55 \\

DPR & 22.78 & 53.44 & 68.58 & 35.92 \\
ANCE & 24.54 & 54.21 & 69.08 & 36.88 \\
RepBERT & 22.57 & 52.20 & 65.65 & 35.13 \\
Sentence-T5 & 22.51 & 52.00 & 65.12 & 34.95 \\

\midrule
\multicolumn{5}{l}{\textit{Generative retrieval}} \\
\midrule
DSI (SI) & 27.42 & 47.26 & 56.58 & 34.31 \\
DSI-QG (SI) & 30.17 & 53.20 & 66.37 & 38.85 \\
NCI (SI) & 32.69 & 55.82 & 69.20 & 42.84 \\
SEAL (NG) & 29.30 & 54.12 & 68.53 & 40.34 \\
Ultron (TU) & 33.78 & 54.20 & 67.05 & 42.51 \\
Ultron (PQ) & 25.64 & 53.09 & 65.75 & 37.12 \\
ROGER-NCI (SI) & 33.20 & 56.34 & 69.80 & 43.45 \\
ROGER-Ultron (TU) & 35.90 & 55.59 &69.86 & 44.92 \\
MINDER (SI) & 31.00 & 55.50 & 65.79 & 43.50 \\
LTRGR (SI) & 32.80 & 56.20 & 68.74 & 44.80 \\

\midrule
\multicolumn{5}{l}{\textit{RL-based Generative retrieval}} \\
\midrule
GenRRL (TU) & 35.79 & 56.49 &\underline{70.96} & 45.73 \\
GenRRL (Sum) & 36.32 & 57.42 & \textbf{71.49} & 46.31 \\
DDRO (TU) & 39.70 & 51.96 & 54.17 & 44.72 \\
DDRO (PQ) &\underline{48.10} &\underline{62.64} &66.84 &\underline{54.32} \\

\midrule
\multicolumn{5}{l}{\textit{Ours}} \\
\midrule
TCA-GRPO (TU) & 40.63\rlap{$^\ddagger$} & 52.42 & 55.39 & 45.53\rlap{$^\ddagger$} \\
TCA-GRPO (PQ) & \textbf{50.02}\rlap{$^\ddagger$} & \textbf{64.09}\rlap{$^\ddagger$} & 67.36 & \textbf{56.10}\rlap{$^\ddagger$} \\
\bottomrule
\end{tabular}
}
\vspace{-4mm}
\end{table}

\subsection{Implementation details}
The SFT model is based on the T5-base pretrained model following DDRO~\cite{DDRO}, and we initialize the policy model from the SFT model and use an identical frozen copy as the reference model. Gold hidden-state trajectories are precomputed offline under teacher forcing. During reinforcement learning, candidate DocIDs are sampled with trie-based constrained decoding to ensure that generated sequences remain within the valid DocID space. We train the policy model for 4 epochs with a learning rate of $1 \times 10^{-6}$ and a batch size of 256. For TCA-GRPO, unless otherwise specified, the group size is set to $G=8$. The KL coefficient is set to $\beta=0$ for PQ-based DocIDs and $\beta=0.01$ for TU-based DocIDs. During evaluation, we use constrained beam search to generate ranked DocID lists. All experiments were conducted on 8 NVIDIA A100 GPUs.

\section{Experimental Results and Analysis}
\subsection{Research questions}
In this section, we present the experimental results to systematically evaluate the performance of TCA and analyze the contribution of its core components.
We list the following research questions to guide our experiments: 
\begin{enumerate*}[label=\textbf{RQ\arabic*}, nosep]
    \item How does our method perform compared with baselines?\label{Q:main_results}
    \item Can the proposed TCA be integrated with different policy optimization algorithms, and how does TCA-GRPO compare with TCA-PPO for generative retrieval?\label{Q:TCA_with_other_RL}
    \item Does token-level credit assignment provide additional benefits over sequence-level reward optimization? \label{Q:token_level_reward}
    \item How do training dynamics and key optimization choices affect retrieval performance?\label{Q:hyperparameter}
    \item How does TCA-GRPO improve DocID generation at the token level in representative retrieval cases?\label{Q:case_study}
\end{enumerate*}

\subsection{Performance comparison (RQ1)}

Tables~\ref{tab:msmarco_main} and~\ref{tab:nq_main} report the overall retrieval results on MS MARCO and Natural Questions datasets. 
Across the two benchmarks, our TCA-based methods show clear gains in Recall performance, with \textsc{TCA-GRPO} delivering the most consistent improvements. 
On MS MARCO dataset, \textsc{TCA-GRPO} with TU DocID obtains the best R@1, R@5, and MRR@10 among all compared methods, reaching 39.60, 68.19, and 51.10, respectively. 
Compared with the reproduced DDRO model under the same TU type DocID setting, it improves R@1 from 38.24 to 39.60 and MRR@10 from 50.33 to 51.10. 
On Natural Questions, \textsc{TCA-GRPO} with PQ DocID achieves the best R@1 and MRR@10, improving DDRO from 48.10 to 50.02 in R@1 and from 54.32 to 56.10 in MRR@10.

These results show that token-level credit assignment is effective for improving generative retrieval. 
Compared with prior generative retriever methods, \textsc{TCA} further improves the alignment between autoregressive token generation and document-level relevance. 
Rather than assigning the same relevance signal to the entire DocID sequence tokens, \textsc{TCA} provides differentiated feedback to intermediate token decisions, which helps the model assign higher probability to trajectories leading to relevant documents.

We also observe that the improvements are more pronounced on R@1 and MRR@10 than on R@5 and R@10. 
For example, on MS MARCO dataset, \textsc{TCA-GRPO} achieves the best R@1 and MRR@10, while GenRRL obtains the highest R@10. 
A similar trend appears on Natural Questions dataset, where GenRRL remains strong on R@10.
This suggests that TCA mainly improves early precision and top-ranking quality, while broader recall may still benefit from document-level or list-level relevance optimization. 
This observation is consistent with our motivation: token-level credit assignment is designed to refine the generation trajectory and improve the ranking of highly relevant DocIDs at the top positions.

\begin{table}[t]
\centering
\caption{Performance comparison between TCA-GRPO and TCA-PPO. 
The best results are shown in bold, and the second-best results are underlined.
}
\label{tab:grpo-vs-ppo}
\resizebox{\columnwidth}{!}{
\begin{tabular}{ll|cccc}
\toprule
Setting & Method & R@1 & R@5 & R@10 & MRR@10 \\

\midrule
\multirow{3}{*}{MS-TU} & baseline-sft & 38.11 & 65.71 & 73.39 & 49.51 \\
 & TCA-PPO &\underline{39.35} &\underline{66.58} &\underline{74.87} & \underline{50.63} \\
 & TCA-GRPO & \textbf{39.60} & \textbf{68.19} & \textbf{75.37} & \textbf{51.10} \\
\midrule
\multirow{3}{*}{MS-PQ} & baseline-sft & 30.19 &\underline{62.50} &\underline{71.16} & 43.57 \\
 & TCA-PPO &\underline{31.68} &\textbf{62.87} &\textbf{72.02} &\underline{44.63} \\
 & TCA-GRPO & \textbf{32.42} &61.88 &\underline{71.16} & \textbf{44.85} \\

\midrule
\multirow{3}{*}{NQ-TU} & baseline-sft & 38.99 &\underline{50.84} &\underline{53.59} & 44.05 \\
 & TCA-PPO &\underline{39.17} &50.05 & 53.35 &\underline{44.07} \\
 & TCA-GRPO & \textbf{40.63} & \textbf{52.42} & \textbf{55.39} & \textbf{45.53} \\
\midrule
\multirow{3}{*}{NQ-PQ} & baseline-sft & 42.26 & 56.85 & 61.12 & 48.45 \\
 & TCA-PPO & \underline{49.27} & \textbf{64.15} &\textbf{67.48} & \underline{55.80} \\
 & TCA-GRPO & \textbf{50.02} & \underline{64.09} & \underline{67.36} & \textbf{56.10} \\

\bottomrule
\end{tabular}
}
\vspace{-3mm}
\end{table}
\subsection{Impact of policy optimization method (RQ2)}
To examine whether the proposed TCA reward design is tied to a particular policy optimizer, we compare two instantiations of TCA: TCA-PPO and TCA-GRPO. As described in Section~\ref{sec:TCA-PPO}, TCA-PPO uses the same token-level reward design as TCA-GRPO but replaces group-relative advantage estimation with value-function-based advantage estimation. This comparison therefore tests the optimizer-level generality of TCA.
As shown in Table~\ref{tab:grpo-vs-ppo}, \textsc{TCA-PPO} improves over \texttt{baseline-sft} in most settings, especially on the recall metrics. For example, on NQ-PQ, \textsc{TCA-PPO} improves R@1 from 42.26 to 49.27 and MRR@10 from 48.45 to 55.80. However, its gains are less consistent than those of \textsc{TCA-GRPO}; on NQ-TU, for instance, the improvements in R@1 and MRR@10 are marginal, while R@5 and R@10 slightly decrease.

In contrast, \textsc{TCA-GRPO} consistently improves R@1 and MRR@10 over \texttt{baseline-sft} across all settings and achieves the best R@1 and MRR@10 in every setting. We hypothesize that PPO is less stable because it must learn a value function over DocID prefix states, where future actions are highly discrete and constrained by the DocID trie. GRPO instead estimates advantages through relative comparisons among candidate DocIDs sampled for the same query, which better matches the ranking-oriented nature of retrieval. These results suggest that TCA is not restricted to a single optimizer, while group-relative advantage estimation is better suited to generative retrieval.

\begin{table}[t]
\centering
\caption{Comparison between sequence-level and token-level rewards under TCA-GRPO on MS MARCO-TU and NQ-PQ. All variants are initialized from \texttt{baseline-sft} model. $\dagger$ denotes our reproduced results. The best results are shown in bold, and the second-best results are underlined.}
\label{tab:reward_ablation_ms_nq}
\begin{tabular}{l|cccc}
\toprule
Method & R@1 & R@5 & R@10 & MRR@10 \\

\midrule
\multicolumn{5}{c}{MS MARCO-TU} \\
\midrule
\texttt{baseline-sft} & 38.11 & 65.71 & 73.39 & 49.51 \\
\midrule
DDRO$^\dagger$ & 38.24 &\underline{67.33} & 74.26 & 50.33 \\
TCA-GRPO\textsubscript{seq} &\underline{39.10} &\textbf{68.19} &\underline{74.87} & \underline{50.86} \\
TCA-GRPO &\textbf{39.60} &\textbf{68.19} &\textbf{75.37} &\textbf{51.10} \\

\midrule
\multicolumn{5}{c}{NQ-PQ} \\
\midrule
baseline-sft & 42.26 & 56.85 & 61.12 & 48.45 \\
\midrule
DDRO$^\dagger$ &48.10 & 62.64 & 66.84 & 54.32 \\
TCA-GRPO\textsubscript{seq} &\underline{48.80} &\underline{63.57} &\underline{66.93} &\underline{55.24} \\
TCA-GRPO & \textbf{50.02} &\textbf{64.09} &\textbf{67.36} &\textbf{56.10} \\
\bottomrule
\end{tabular}
\end{table}
\subsection{Token-level vs. sequence-level reward (RQ3)}

To isolate the effect of reward granularity, we compare the proposed token-level trajectory reward with a standard sequence-level reward while keeping the optimizer fixed to GRPO.
As shown in Table~\ref{tab:reward_ablation_ms_nq}, the sequence-level reward already improves over \texttt{baseline-sft}, indicating that reinforcement learning post-training itself is beneficial for generative retrieval.
However, replacing this coarse sequence-level feedback with token-level credit assignment yields consistent additional gains.
On MS MARCO-TU, the token-level reward improves R@1 from 39.10 to 39.60 and MRR@10 from 50.86 to 51.10 compared with the sequence-level reward.
On NQ-PQ, the gains are larger, with R@1 increasing from 48.80 to 50.02 and MRR@10 from 55.24 to 56.10.
These results suggest that assigning the same terminal reward to every generated DocID token provides useful but insufficient supervision.
By providing position-wise feedback, the token-level reward more precisely reinforces decisions that keep the generation trajectory close to the gold DocID.

\begin{figure}[t]
    \centering
    \includegraphics[width=0.82\linewidth]{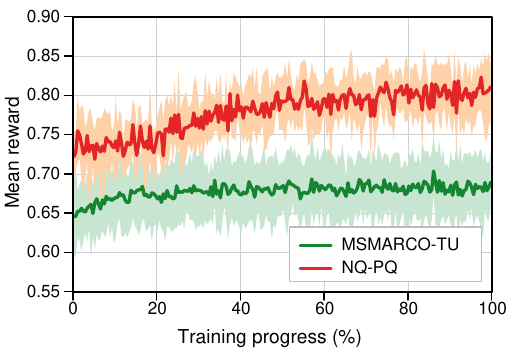}
    \caption{
    Reward dynamics of \textsc{TCA-GRPO} on MS MARCO-TU and NQ-PQ.
    Training steps are grouped into equal-width progress bins.
    Bold curves show the mean per-step reward in each bin, and shaded bands indicate the 10th--90th percentile range.
    }
    \label{fig:reward_dynamics}
    \vspace{-4mm}
\end{figure}

\subsection{Rewards during TCA-GRPO training (RQ4)}
Figure~\ref{fig:reward_dynamics} illustrates the evolution of reward values during the training process using TCA-GRPO.
We plot reward trajectories for two representative settings, MS MARCO-TU and NQ-PQ.
Because per-step rewards fluctuate substantially during policy optimization, we group training steps into equal-width progress bins and report both the mean reward and the 10th--90th percentile range within each bin.
The mean reward increases consistently in both settings, indicating that \textsc{TCA-GRPO} can progressively achieve higher rewards under different datasets and DocID forms.
However, the pace of improvement differs across settings.
On NQ-PQ, the reward rises quickly during approximately the first 40\% of training, followed by a slower but steady increase thereafter, whereas on MS MARCO-TU the reward improves more gradually throughout training.
We attribute this difference to the data and identifier characteristics of the two settings: NQ-PQ is built on a smaller and relatively cleaner benchmark with compact semantic-code identifiers, which may make reward-driven trajectory refinement easier; in contrast, MS MARCO-TU contains noisier web documents and lexical title-URL identifiers, leading to a larger and more heterogeneous generation space.
Despite these different convergence patterns, the rewards in both settings maintain a stable upward trend, suggesting that TCA-GRPO provides a robust training signal for model optimization.

\subsection{Training strategy and hyperparameter (RQ4)}

\textbf{Constrained decoding.}
\begin{table}[t]
\centering
\caption{Effect of constrained decoding during TCA-GRPO rollout on NQ-PQ and MSMARCO-TU. The best results are shown in bold.}
\label{tab:train_constraints}
\begin{tabular}{l|cccc}
\toprule
Method & R@1 & R@5 & R@10 & MRR@10 \\
\midrule
\multicolumn{5}{l}{NQ-PQ} \\
w/o constraints & 45.18 & 59.77 & 63.28 & 51.40 \\
w/ constraints &\textbf{50.02} &\textbf{64.09} &\textbf{67.36} &\textbf{56.10} \\

\midrule
\multicolumn{5}{l}{MSMARCO-TU} \\
w/o constraints & 37.74 & 65.96 & 75.12 & 49.60 \\
w/ constraints &\textbf{39.60} &\textbf{68.19} &\textbf{75.37} &\textbf{51.10} \\
\bottomrule
\end{tabular}
\end{table}
Constrained decoding during rollout is important for stable RL training. 
As shown in Table~\ref{tab:train_constraints}, removing constraints substantially hurts performance. 
On NQ-PQ, constrained rollout improves R@1 from 45.18 to 50.02 and MRR@10 from 51.40 to 56.10. 
This indicates that unconstrained sampling often produces invalid or low-quality DocID trajectories, making the trajectory reward noisy. 
Restricting rollouts to the legal DocID space makes sampled candidates more meaningful and improves token-level credit assignment.

\begin{figure}[t]
    \centering
    \includegraphics[width=0.8\columnwidth]{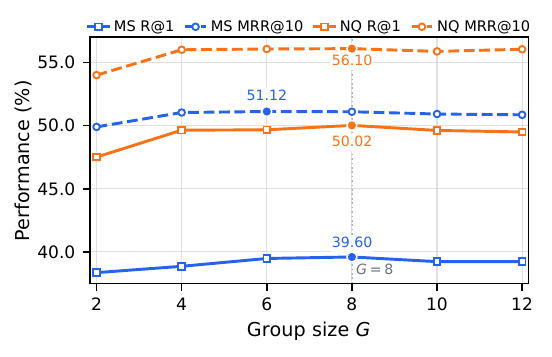}
    \caption{Effect of group size in TCA-GRPO on MS MARCO-TU and NQ-PQ.}
    \label{fig:group_size}
\end{figure}

\begin{figure}[t]
    \centering
    \includegraphics[width=0.8\linewidth]{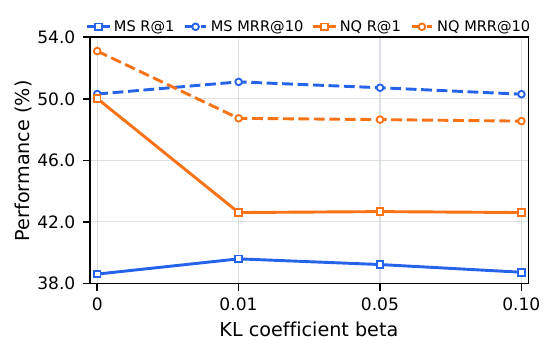}
    \caption{
    Effect of the KL regularization coefficient $\beta$ on R@1 and MRR@10 for MS-TU and NQ-PQ under \textsc{TCA-GRPO}.
    }
    \label{fig:beta_r1_mrr}
\end{figure}

\textbf{Group size.} Figure~\ref{fig:group_size} shows that increasing the group size from $G=2$ to moderate values consistently improves R@1 and MRR@10 on both MS MARCO-TU and NQ-PQ. 
The performance stabilizes when $G$ ranges from 4 to 8, with $G=8$ achieving the best or near-best results in terms of R@1 and MRR@10. 
Further increasing the group size does not yield consistent gains, suggesting that a moderate number of sampled candidates is sufficient for reliable group-relative advantage estimation.

\textbf{KL coefficient.} Figure~\ref{fig:beta_r1_mrr} shows the effect of the KL regularization coefficient on two representative metrics, R@1 and MRR@10.
The preferred KL strength differs between the TU and PQ settings.
For MS-TU, a small KL coefficient, $\beta=0.01$, yields the best performance on both R@1 and MRR@10, suggesting that a mild constraint to the SFT reference policy helps preserve the lexical generation patterns of title- and URL-based identifiers.
In contrast, for NQ-PQ, removing KL regularization performs best: $\beta=0$ achieves the highest R@1 and MRR@10, whereas non-zero KL coefficients lead to clear performance degradation.
This result suggests that PQ DocIDs, which consist of discrete semantic code tokens, are more sensitive to KL constraints during reward-driven policy optimization than lexical TU identifiers.
Based on these observations, we set $\beta=0.01$ for TU DocIDs and $\beta=0$ for PQ DocIDs in the main experiments.

\subsection{Case study (RQ5)}

\begin{figure}[t]
    \centering
    \includegraphics[width=0.85\linewidth]{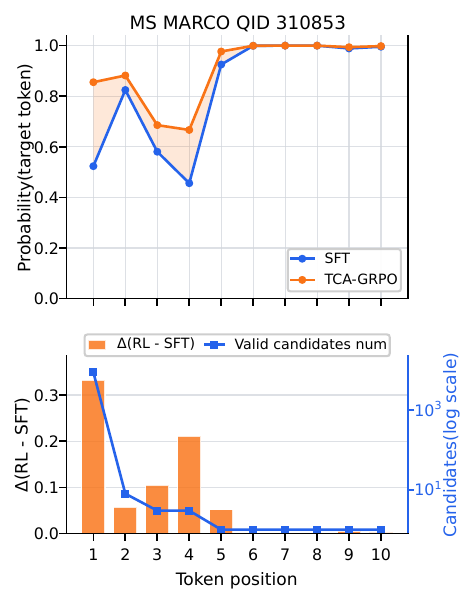}
    \caption{
    Case study on MS MARCO for the query ``how much does a pharmacy tech make?''. 
    The upper panel compares the probabilities assigned by the SFT model and \textsc{TCA-GRPO} to the gold next token under the gold DocID prefix. 
    The lower panel shows the probability differences after TCA-GRPO and the number of valid DocID candidates compatible with the gold prefix at each decoding position.
    }
    \label{fig:case_study}
\end{figure}

To qualitatively examine how TCA improves DocID generation, we analyze a representative MS MARCO-TU query: ``how much does a pharmacy tech make?'' The gold document is \texttt{D702099}, whose TU DocID is ``pharmacy technician salaries www.glassdoor.com''.
With constrained beam-search inference using 10 beams, the SFT model ranks the DocID only at the 9th position, whereas \textsc{TCA-GRPO} promotes it to rank 1. This result shows that token-level credit assignment can translate into a direct improvement in the final retrieval ranking.
Figure~\ref{fig:case_study} explains this improvement from the token-generation perspective. Under the target DocID prefix, \textsc{TCA-GRPO} assigns higher probabilities to the correct next token at 9 out of 10 decoding positions, increasing the average gold-token probability from 0.829 to 0.906. 
The gains are especially pronounced at early and middle positions: the probability of ``pharmacy'' at the first position increases from 0.523 to 0.855, ``salaries'' at the third position increases from 0.580 to 0.686, and ``www'' at the fourth position increases from 0.456 to 0.667.
These improvements are particularly important because early decoding steps correspond to a much larger candidate space. At the first token position, the gold prefix is compatible with 8,980 valid DocID continuations. After generating the first token, the number of compatible candidates rapidly drops to 8, and further decreases to 3 after the second token. Once the path becomes nearly determined, later gold tokens already receive probabilities close to 1.0 under SFT, leaving limited room for further improvement. Therefore, TCA mainly strengthens the gold trajectory at the more uncertain and influential decoding positions, which increases the sequence score of the relevant DocID and helps move it from rank 9 to rank 1.
\section{Conclusion}
Generative retrieval generates DocIDs token by token, but existing RL feedback is usually document-level. This paper addresses the mismatch with \textsc{TCA}, which compares generated hidden states with gold DocID trajectories from a frozen reference model to assign token-level trajectory rewards.
\textsc{TCA} improves generative retrieval by turning document-level supervision into dense feedback over the actual decoding trajectory. Across MS MARCO and NQ, \textsc{TCA-GRPO} consistently improves top-ranking metrics over supervised fine-tuning and sequence-level reinforcement learning baselines, while \textsc{TCA-PPO} shows that the reward design is not tied to a single optimizer. The reward ablations further indicate that the gains come from token-level credit assignment rather than reinforcement learning alone, and the training analyses show that constrained rollout, moderate group size, and DocID-dependent KL regularization are important for stable and effective optimization.
These results also highlight two practical limitations. First, \textsc{TCA} introduces a more complex training pipeline because it relies on a supervised reference model before reinforcement learning optimization. Second, the reinforcement learning stage adds nontrivial training overhead due to rollout generation and token-level reward computation. Simplifying this two-stage procedure and improving the efficiency of RL-based optimization remain important directions for future work. 
\section{Acknowledgments}
This work was supported by the Shandong Provincial Natural Science Foundation (ZR2022QF004, ZR2021QF129), the National Natural Science Foundation of China (62202271, 61902219, 61972234, 61672324, 62072279, 62102234, 62272274), the National Key R\&D Program of China (2020YFB1406704, 2022YFC3303004), the Key Scientific and Technological Innovation Program of Shandong Province with grants No. 2025CXGC010108 and No. 2019JZZY010129, the Tencent WeChat Rhino-Bird Focused Research Program (WXG-FR-2023-07).
All content represents the opinion of the authors, which is not necessarily shared or endorsed by their respective employers and/or sponsors.

\newpage
\section{GenAI Usage Disclosure}
This work involved the use of generative AI tools in limited and supervised ways. Specifically, docT5query was used to assist with generating pseudo-queries following prior works, during the data preparation stage. In addition, OpenAI's ChatGPT was employed solely for minor language polishing, such as grammar correction and wording improvements. All uses of GenAI tools were conducted under the full supervision of the authors, who are solely responsible for the final content.

\bibliographystyle{ACM-Reference-Format}
\bibliography{reference}
\end{document}